\documentclass[journal]{IEEEtran}

\usepackage{xcolor,soul,framed} %,caption

\colorlet{shadecolor}{yellow}
\usepackage[pdftex]{graphicx}
\graphicspath{{../pdf/}{../jpeg/}}
\DeclareGraphicsExtensions{.pdf,.jpeg,.png}
\usepackage{amsmath} 

\usepackage{array}
\usepackage{mdwmath}
\usepackage{mdwtab}
\usepackage{eqparbox}
\usepackage{url}
\usepackage{algorithm}
\usepackage{algpseudocode}
\usepackage{amsmath,amssymb,bm}
\usepackage{caption}
\usepackage{algpseudocode}

\begin{document}
\bstctlcite{IEEEexample:BSTcontrol}
    \title{Efficient MPC for LC-Filtered Three-Level NPC Inverters in Vehicle-to-Load Applications}
    
\author{%
\begin{tabular}{c@{\hspace{0.1cm}}c}
\begin{minipage}[t]{0.5\textwidth}\centering
Rock Agon \\
\textit{Department of Electrical and Computer Engineering} \\
\textit{Carnegie Mellon University Africa} \\
Kigali, Rwanda \\
ragon@alumni.cmu.edu
\end{minipage}
&
\begin{minipage}[t]{0.5\textwidth}\centering
Dieudonne Iradukunda \\
\textit{Department of Electrical and Computer Engineering} \\
\textit{Carnegie Mellon University Africa} \\
Kigali, Rwanda \\
dieudoni@alumni.cmu.edu
\end{minipage}
\end{tabular}%
}


% ====================================================================
\maketitle

% === ABSTRACT ====================================================================
% =================================================================================
\begin{abstract}
Finite control-set model predictive control (FCS-MPC) offers fast, modulator-free voltage regulation but is sensitive to the fidelity of its internal model. This sensitivity is usually countered by algorithmic augmentation, disturbance observers, adaptive weighting, or learning-based estimators, at the cost of extra tuning and computation. This paper instead improves fidelity at the source. For a three-level neutral-point-clamped (3L-NPC) inverter with an LC output filter in vehicle-to-load operation, a coupled alphabeta-frame predictive model carries the capacitor voltage and filter current jointly through the state vector, so that the second-order LC coupling is retained without enlarging the 27-candidate search. The cost function additionally regulates dc-link neutral-point balance and switching effort. One-step prediction accuracy, parametric sensitivity, and computational cost are bounded in closed form. Simulation of a V2L system yields a voltage total harmonic distortion of 0.72\%, a steady-state tracking error of 1.6\%, and transient recovery within one fundamental cycle, meeting the applicable IEEE power-quality limits with margin, without observers, parameter estimation, or learning-based augmentation.
\end{abstract}

% === KEYWORDS ====================================================================
% =================================================================================
\begin{IEEEkeywords}
electric vehicle-to-load, model predictive control, three-level neutral-point-clamped inverter, voltage control.
\end{IEEEkeywords}

\IEEEpeerreviewmaketitle

% ====================================================================

% === I. INTRODUCTION ==============================================

\section{Introduction}
Multilevel power converters, particularly the three-level Neutral-Point-Clamped (3L-NPC) inverters, have attracted significant attention in voltage regulation applications due to their inherent advantages [1]. These advantages include reduced total harmonic distortion, improved efficiency, lower switching stress, and reduced dv/dt stress compared to conventional two-level converters [2]-[4]. However, the increased number of switching states and the presence of internal dynamics, such as neutral-point voltage balancing, significantly increase control complexity. Therefore, achieving accurate voltage control often requires additional circuitry or specialized balancing strategies embedded within the modulation scheme [5], thereby increasing the design complexity and computation.\\

Conventional control approaches typically employ cascaded proportional–integral (PI) regulators combined with pulse-width modulation (PWM) [6]. Although widely adopted in industrial practice, these methods are inherently limited by control bandwidth and loop interactions, which can restrict performance under fast varying operating conditions [7], [8]. To overcome these limitations, predictive control strategies have gained considerable attention. Early approaches, such as deadbeat control has been widely applied to the control of inverters, rectifiers, active power filters, power-factor preregulators, and UPS systems [9]. It offers fast transient response but suffer from sensitivity to parameter uncertainties and measurement noise [10]. Later, Model Predictive Control (MPC) emerged as a more general and flexible predictive control framework capable of addressing many of these limitations [11]. In MPC, the future evolution of the system states is predicted over a finite prediction horizon. At each sampling instant, an optimization problem is solved to minimize a predefined cost function while satisfying the constraints of the system [12]. Only the first control action of the optimal sequence is then applied to the system, and the optimization process is repeated at the next sampling instant according to a receding-horizon strategy. This makes MPC particularly attractive for power electronic converters and drives [13].\\

MPC implementations are typically categorized into continuous-control-set MPC (CCS-MPC) and finite-control-set MPC (FCS-MPC). In CCS-MPC, a modulator is required to generate the switching signals with a fixed switching frequency. In contrast, FCS-MPC directly exploits the discrete nature of the power electronic converters. Since voltage-source inverters can generate only a finite set of switching states, the MPC optimization problem can be simplified by evaluating the predicted system behavior exclusively for these admissible switching states [14]. This approach eliminates the need for an external modulator and integrates the control and switching decisions into a single optimization process based on the cost function [15]-[19]. Consequently, FCS-MPC provides fast dynamic response and a conceptually simple control structure, which has contributed to its widespread adoption in power electronics applications [20], [21]. This approach has been extensively applied to various converter topologies, including three-phase active rectifiers [22],[23], four-leg inverters [24], grid-connected converters [25], and permanent magnet synchronous motor drives [26]-[28].\\

Despite these advantages, the performance of FCS-MPC is
fundamentally dependent on the accuracy of the predictive model. This limitation is especially critical in LC-filtered inverters, where strong coupling between filter current and output voltage governs system dynamics [16]. Inaccurate representation of this coupling leads to degraded voltage regulation and increased harmonic distortion. Various approaches have been proposed to mitigate these
issues, including observer-based methods, adaptive weighting strategies, and more recently data-driven techniques including neural networks, reinforcement learning, and metaheuristic optimization, have been integrated with MPC [29]–[32]. While effective in certain cases, these methods often introduce additional tuning complexity, sensitivity to noise, or increased computational burden, limiting their suitability for real-time embedded applications.\\

In contrast, this paper investigates the extent to which predictive performance can be improved through model fidelity while preserving the simplicity of conventional FCS-MPC structures. Rather than introducing observers, adaptive weighting methods, or machine-learning-assisted optimisation, the proposed controller employs a coupled predictive model that retains both filter current and output voltage states within the prediction horizon. The objective is not to increase algorithmic sophistication, but to examine how far careful physical modelling alone can improve voltage regulation performance in LC-filtered 3L-NPC inverters. By incorporating these coupled dynamics into the prediction model, the proposed approach improves voltage tracking accuracy and robustness to load variations without increasing algorithmic complexity.  The proposed modeling framework enhances control performance without requiring algorithmic augmentation, while maintaining computational efficiency and preserving clear physical interpretability. The main contributions of this paper are as follows:

\begin{itemize}
\item An investigation into the performance gains achievable through predictive model fidelity without additional algorithmic augmentation.

\item An FCS-MPVC formulation for LC-filtered 3L-NPC inverters that retains the complete filter-state representation while preserving the computational simplicity of conventional one-step predictive control.

\item A closed-form analytical characterisation of the one-step
        prediction error, covering disturbance-truncation, dc-link
        voltage, discretisation, and parametric contributions, together
        with a parametric-sensitivity bound.
\end{itemize}
Since the superiority of MPC over traditional control methods has
already been extensively demonstrated in the literature, such
comparisons are outside the scope of this work.

The rest of this paper is organized as follows. Section 2 introduces the V2L concept. Section 3 describes the overall system architecture. Section 4 presents the proposed algorithm. Section 5 details the simulation setup and results. Section 6 discusses the findings, and Section 7 concludes.
 
\section{System Modelling}
\label{sec:system_modelling}
 
V2L technology enables an electric
vehicle (EV) to supply power to standalone loads in the absence
of a grid connection, which makes it well suited for emergency and
portable power applications. In this operating mode the
power-electronic interface behaves as a controlled voltage
source delivering a stable magnitude and frequency under varying
load conditions.
 
\begin{figure}[H]
  \begin{center}
  %\centering
  \includegraphics[width=\columnwidth]{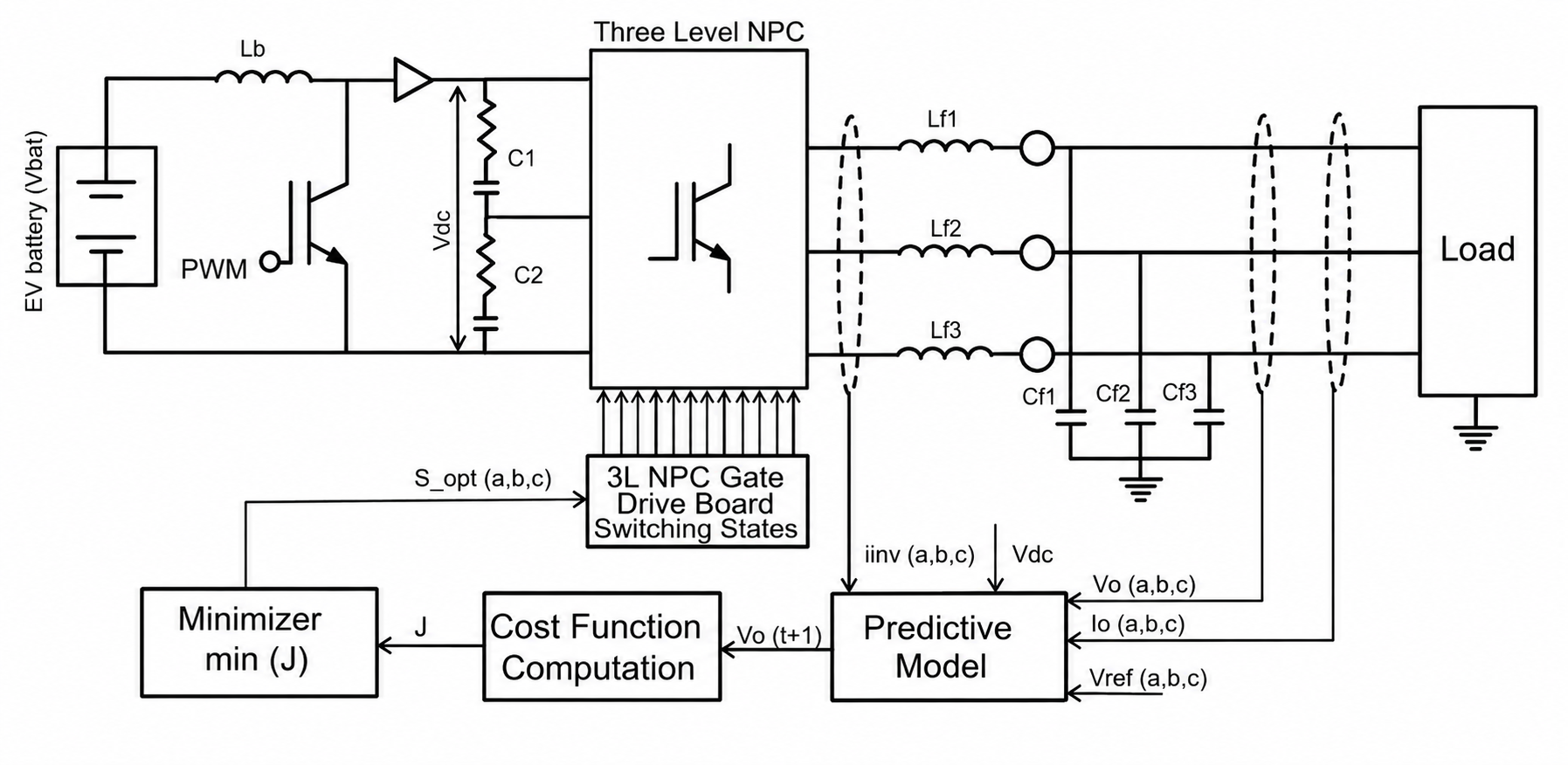}
  \caption{V2L System}
  \label{fig:v2l_system}
  \end{center}
\end{figure}

The V2L system considered in this work, shown in
Fig.~\ref{fig:v2l_system}, comprises: (i) an EV battery, (ii) a
unidirectional dc--dc boost converter that regulates the dc-link
voltage, (iii) a 3L-NPC inverter,
(iv) a three-phase $LC$ output filter, and (v) a load.

\subsection{DC-Link Boost Converter}
 
The boost converter regulates the dc-link voltage $V_{dc}$ from the
battery voltage $V_{\mathrm{bat}}$. Its dynamics are defined by:
\begin{align}
L_b\,\frac{d i_L}{dt} &= V_{\mathrm{bat}} - (1-D)\,V_{dc}
\label{eq:boost_iL}\\
C_{dc}\,\frac{d V_{dc}}{dt} &= (1-D)\,i_L - i_{\mathrm{inv}}
\label{eq:boost_vdc}
\end{align}
where $L_b$ and $C_{dc}$ denote the boost inductance and dc-link
capacitance, $i_L$ is the boost-inductor current,
$i_{\mathrm{inv}}$ is the dc-side current drawn by the inverter, and
$D\in[0,1]$ is the duty cycle of the converter switch. The boost
stage is regulated through an outer dc-link voltage loop. 
 
\subsection{Three-Level NPC Inverter}
\label{sec:npc_inverter}

\begin{figure}[H]
  %\begin{center}
  \centering
  \includegraphics[width=\columnwidth]{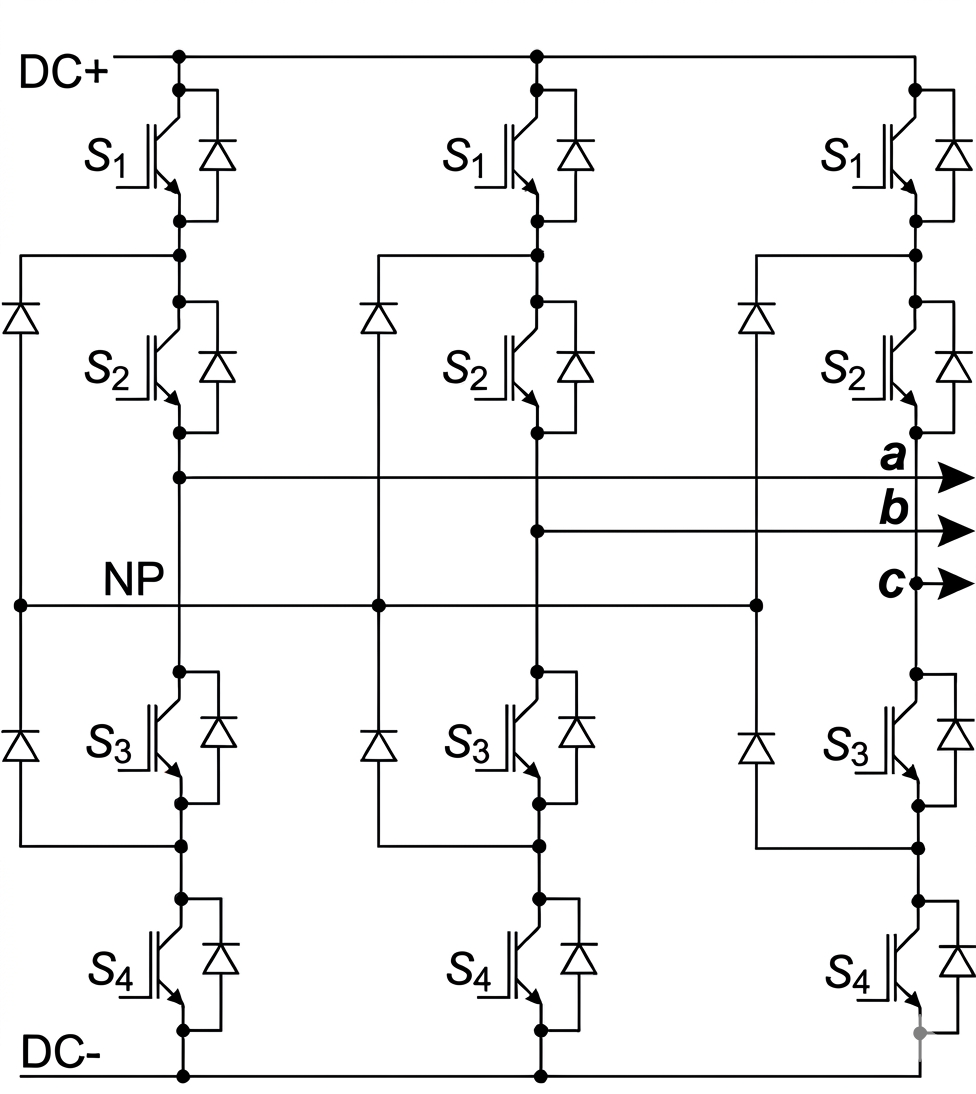}
  \caption{Three-level neutral-point-clamped}
  \label{fig:npc_topology}
  %\end{center}
\end{figure}

The 3L-NPC inverter is shown
in Fig.~\ref{fig:npc_topology}.

Each phase leg consists of four active switches and two clamping diodes, and the dc-link is
formed by two series-connected capacitors $C_1$ and $C_2$ that
define a midpoint $N$ used as the neutral reference for the
output. The complementary structure of the switches prevents
short-circuit paths through the dc-link and reduces the four
binary switch states to three admissible leg states, summarized
in Table~\ref{tab:switching_states}.

\begin{table}[H]
\centering
\caption{Switching States and Phase Voltages of a 3L-NPC}
\label{tab:switching_states}
\renewcommand{\arraystretch}{1.2}
\begin{tabular}{c c c c c c}
$S_x$ & $S_{x1}$ & $S_{x2}$ & $S_{x3}$ & $S_{x4}$ & $v_{xN}$ \\
\hline

$+1$ & on  & on  & off & off & $+V_{dc}/2$ \\
\hline
$\phantom{+}0$  & off & on  & on  & off & $0$ \\
\hline
$-1$ & off & off & on  & on  & $-V_{dc}/2$ \\
\hline

\end{tabular}
\end{table}

The three values of $S_x$ correspond to the upper rail ($+1$),
the dc-link midpoint ($0$), and the lower rail ($-1$), giving
leg-to-midpoint voltages of $+V_{dc}/2$, $0$, and $-V_{dc}/2$,
respectively. 

The switching state of phase
$x\in\{a,b,c\}$ is denoted
\begin{equation}
S_x \in \{-1,\,0,\,+1\},
\qquad
S = [\,S_a,\;S_b,\;S_c\,]^{\top}
\label{eq:Sx-Svec}
\end{equation}

and since each phase admits three values, the inverter has 27 admissible switching combinations, which generate a
hexagonal space-vector pattern characteristic of the 3L-NPC. The leg-to-midpoint voltage is: 
\begin{equation}
v_{xN} \;=\; S_x\,\frac{V_{dc}}{2}, \qquad x \in \{a,b,c\}.
\label{eq:leg_voltage}
\end{equation}

The leg-to-midpoint voltages of \eqref{eq:leg_voltage} are
referenced to the dc-link midpoint $N$, but the load sees
voltages referenced to its own neutral $n$. For a three-wire
load, no current returns through $N$, so the load neutral
floats with respect to $N$ at a potential $v_{xN}$ defined as: 

\begin{equation}
v_{nN} = \frac{1}{3}\left[v_{aN}+v_{bN}+v_{cN}\right]
\label{eq:vnN}
\end{equation}
and the phase-to-load-neutral voltages are obtained by
subtracting this common potential from each leg voltage:
\begin{equation}
\mathbf{v}_i \;=\;
\begin{bmatrix} v_{an}\\ v_{bn}\\ v_{cn} \end{bmatrix}
\;=\;
\frac{V_{dc}}{2}\,\mathbf{T}\,\mathbf{S},\quad
\mathbf{T} \;=\; \frac{1}{3}\!
\begin{bmatrix}
\phantom{-}2 & -1 & -1 \\
-1 & \phantom{-}2 & -1 \\
-1 & -1 & \phantom{-}2
\end{bmatrix}
\label{eq:vi_abc}
\end{equation}
The matrix $\mathbf{T}$ performs the change of reference from
the dc-link midpoint to the load neutral.

For the predictive controller,
the three-phase quantities are projected onto the stationary
$\alpha\beta$ frame using Clarke
transformation: 
\begin{equation}
\mathbf{v}_i{\alpha\beta} \;=\; \frac{2}{3}\!
\begin{bmatrix}
1 & -\tfrac{1}{2} & -\tfrac{1}{2} \\[2pt]
0 & \tfrac{\sqrt{3}}{2} & -\tfrac{\sqrt{3}}{2}
\end{bmatrix}
\mathbf{v}_i
\label{eq:clark}
\end{equation}
Substituting~\eqref{eq:vi_abc} into~\eqref{eq:clark} gives the
direct switching-state-to-$\alpha\beta$-voltage mapping:
\begin{equation}
\mathbf{v}_i{\alpha\beta} \;=\; \frac{V_{dc}}{2}\,
\mathbf{K}\,\mathbf{S},
\label{eq:vi_alphabeta}
\end{equation}
where $\mathbf{K}$ is the constant matrix obtained from the
product of the Clarke and common-mode-removal matrices.
Equation~\eqref{eq:vi_alphabeta} produces a finite set of $27$
$\alpha\beta$ voltage vectors that the predictive controller
selects from.

\subsection{LC Output Filter}
\label{sec:lc_filter}

The inverter is connected to the load through a three-phase
$LC$ filter. The filter attenuates the switching-frequency harmonics produced by the
discrete switching action of the inverter, so that the voltage
delivered to the load is essentially sinusoidal. Additionally, the
inductance $L_f$ and capacitance $C_f$ together define the
second-order dynamics that the predictive controller must
respect when selecting switching states.

Let $\mathbf{v}_o$ denote the three-phase capacitor voltage at
the load terminals, $\mathbf{i}_f$ the inverter-side filter
current flowing through $L_f$, and $\mathbf{i}_o$ the load
current. Applying Kirchhoff's voltage and current laws yields the two
governing equations:
\begin{align}
L_f\,\frac{d \mathbf{i}_f}{dt} &= \mathbf{v}_i - \mathbf{v}_o,
\label{eq:ind_dyn_abc}\\
C_f\,\frac{d \mathbf{v}_o}{dt} &= \mathbf{i}_f - \mathbf{i}_o.
\label{eq:cap_dyn_abc}
\end{align}

Two physical observations follow from
\eqref{eq:ind_dyn_abc}--\eqref{eq:cap_dyn_abc} and are central to the controller design:

\emph{First}, the two equations are bidirectionally coupled.
The output voltage $\mathbf{v}_o$ appears in the inductor
equation and therefore influences the evolution of
$\mathbf{i}_f$, while the filter current $\mathbf{i}_f$ appears
in the capacitor equation and therefore influences the
evolution of $\mathbf{v}_o$. The $LC$ filter is consequently a
single second-order system with two state variables, not two
independent first-order systems. The predictive controller
proposed in Section~\ref{sec:fcs_mpvc} will retain both states and
preserves this coupling.

\emph{Second}, the load current $\mathbf{i}_o$ appears only in
the capacitor equation \eqref{eq:cap_dyn_abc} and acts as an
exogenous disturbance on the output-voltage dynamics. It is
not influenced by the controller and is therefore treated as a
measurable disturbance in the formulation that follows.

For control design, equations
\eqref{eq:ind_dyn_abc}--\eqref{eq:cap_dyn_abc} are projected
onto the stationary $\alpha\beta$ frame. Under balanced three-phase operation, the
$LC$ filter is decoupled across the $\alpha$ and $\beta$ axes,
so that each axis sees an identical scalar second-order
subsystem. The filter dynamics are:
\begin{align}
L_f\,\frac{d \mathbf{i}_f{\alpha\beta}}{dt} &=
\mathbf{v}_i{\alpha\beta} - \mathbf{v}_o{\alpha\beta},
\label{eq:ind_dyn}\\
C_f\,\frac{d \mathbf{v}_o{\alpha\beta}}{dt} &=
\mathbf{i}_f{\alpha\beta} - \mathbf{i}_o{\alpha\beta},
\label{eq:cap_dyn}
\end{align}

% =====================================================================
\section{Finite Set Predictive Voltage Control}
\label{sec:fcs_mpvc}
 
The proposed Finite Control Set Model Predictive Voltage Control (FCS-MPVC) evaluates, at every sampling instant $kT_s$, all admissible switching states of the 3L-NPC inverter and applies the one that minimizes a cost function.
 
\subsection{Continuous-Time State-Space Model}
 
The state vector and control input is:
\begin{equation}
\mathbf{x} \;=\;
\begin{bmatrix}
\mathbf{v}_o{\alpha\beta} \\[2pt] \mathbf{i}_f{\alpha\beta}
\end{bmatrix}
\;\in\; \mathbb{R}^{4}
\label{eq:state_vec}
\end{equation}
The filter dynamics can be written in state-space form:
\begin{equation}
\dot{\mathbf{x}} \;=\;
\mathbf{A}\,\mathbf{x} \;+\;
\mathbf{B}\,\mathbf{\mathbf{v}_i{\alpha\beta} } \;+\;
\mathbf{E}\,\mathbf{i}_o{\alpha\beta},
\label{eq:cont_ss}
\end{equation}
with
\begin{equation}
\mathbf{A} \!=\!\!
\begin{bmatrix}
\mathbf{0}_{2} & \tfrac{1}{C_f}\,\mathbf{I}_{2} \\[2pt]
-\tfrac{1}{L_f}\,\mathbf{I}_{2} & \mathbf{0}_{2}
\end{bmatrix}\!,\;\;
\mathbf{B} \!=\!\!
\begin{bmatrix}
\mathbf{0}_{2} \\[2pt] \tfrac{1}{L_f}\,\mathbf{I}_{2}
\end{bmatrix}\!,\;\;
\mathbf{E} \!=\!\!
\begin{bmatrix}
-\tfrac{1}{C_f}\,\mathbf{I}_{2} \\[2pt] \mathbf{0}_{2}
\end{bmatrix}\!,
\label{eq:ABE}
\end{equation}
where $\mathbf{I}_{2}$ and $\mathbf{0}_{2}$ are the $2\!\times\!2$
identity and zero matrices. 
 
The coupled formulation preserves the complete second-order LC dynamics within the prediction model while maintaining the same candidate switching set and optimization structure as the conventional one-step FCS-MPC. Consequently, the improved model fidelity is achieved without increasing optimization complexity or enlarging the search space.
 
\subsection{Discrete-Time Predictive Model}
 
Since the controller operates in discrete time,
\eqref{eq:cont_ss} is discretised at the sampling period $T_s$
using a zero-order hold Euler formulation on the inverter voltage, giving:
\begin{equation}
\mathbf{x}(k+1) \;=\;
\mathbf{Q}\,\mathbf{x}(k) \;+\;
\mathbf{J}\,\mathbf{\mathbf{v}_o{\alpha\beta} }(k) \;+\;
\mathbf{G}\,\mathbf{i}_o{\alpha\beta}
\label{eq:disc_ss}
\end{equation}
with the transition matrices expressed as:
\begin{equation}
\mathbf{Q} \!=\! e^{\mathbf{A}T_s},\;\;
\mathbf{J} \!=\! \mathbf{A}^{-1}(\mathbf{Q}-\mathbf{I})\mathbf{B},\;\;
\mathbf{G} \!=\! \mathbf{A}^{-1}(\mathbf{Q}-\mathbf{I})\mathbf{E}
\label{eq:zoh_matrices}
\end{equation}

\subsection{One-Step-Ahead Voltage Prediction}
 
For each candidate switching vector $\mathbf{S}^{(m)}$,
$m=1,\dots,N_s$, the corresponding discrete-time predictive model in $\alpha\beta$ is:
\begin{equation}
\mathbf{x}^{(m)}(k+1) \;=\;
\mathbf{Q}\,\mathbf{x}(k) \;+\;
\mathbf{J}\,\mathbf{v}_o{\alpha\beta}^{(m)} \;+\;
\mathbf{G}\,\mathbf{i}_o{\alpha\beta}(k)
\label{eq:state_pred}
\end{equation}
The predicted output voltage is extracted from
\eqref{eq:state_pred} through the output matrix
$\mathbf{C}_v = [\,\mathbf{I}_{2}\;\;\mathbf{0}_{2}\,]$ as follows:
\begin{equation}
\hat{\mathbf{v}}_o{\alpha\beta}^{(m)}(k+1) \;=\;
\mathbf{C}_v\,\mathbf{x}^{(m)}(k+1)
\label{eq:vo_pred}
\end{equation}

\subsection{Neutral-Point Voltage Dynamics}
The two dc-link capacitors $C_1$ and $C_2$ share the midpoint $N$.
Their difference voltage
\begin{equation}
  v_{np} = v_{C_1} - v_{C_2}
  \label{eq:vnp-def}
\end{equation}
is not captured by the filter state and must be
predicted separately, because it is driven by the switching state
through the current injected into the midpoint. A phase leg loads the
midpoint only when it is in the ``0'' state; the instantaneous
neutral-point current is therefore
\begin{equation}
  i_{np}(k) = \sum_{x\in\{a,b,c\}} \big[\, S_x(k)=0 \,\big]\, i_{x}(k),
  \label{eq:inp}
\end{equation}
where $[\,\cdot\,]$ is the Iverson bracket (unity when the leg is
clamped to $N$, zero otherwise) and $i_x$ is the corresponding phase
current. Applying charge balance to the split dc-link over one sampling
interval gives the first-order predictive update
\begin{equation}
  v_{np}(k+1) = v_{np}(k) + \frac{T_s}{C_{dc}}\, i_{np}(k),
  \label{eq:vnp-pred}
\end{equation}
with $C_{dc}=C_1=C_2$. Equation \eqref{eq:vnp-pred} is evaluated for
each candidate switching vector $\mathbf{S}^{(m)}$ alongside the filter
prediction, and adds only a scalar update per candidate, so the
optimisation still operates on the same 27-element admissible set
without additional matrix operations.
 
% ---------------------------------------------------------------------
% NOTE: If the update you actually simulated differs (e.g. you used the
% per-capacitor currents, or a (1-D) split-current form), replace
% \eqref{eq:inp}-\eqref{eq:vnp-pred} with YOUR expression. The text
% above must match the equation you implemented.
% ---------------------------------------------------------------------
 \subsection{Cost Function and Optimal Switching State}
Let $\mathbf{v}^{\star}_{\alpha\beta}(k+1)$ denote the reference output
voltage in the $\alpha\beta$ frame, obtained from a balanced sinusoidal
three-phase reference. Three objectives are combined in the cost
function: output-voltage tracking, and neutral-point balance.
 
The voltage-tracking term is
\begin{equation}
  g^{(m)}_{v} =
  \big\| \mathbf{v}^{\star}_{\alpha\beta}(k+1)
        - \hat{\mathbf{v}}^{(m)}_{o,\alpha\beta}(k+1) \big\|_2^{2}.
  \label{eq:gv}
\end{equation}
The neutral-point term penalises the predicted midpoint imbalance
obtained from \eqref{eq:vnp-pred},
\begin{equation}
  g^{(m)}_{np} = \big( v^{(m)}_{np}(k+1) \big)^{2}.
  \label{eq:gnp}
\end{equation}

The composite cost is the weighted sum
\begin{equation}
  g^{(m)} = g^{(m)}_{v}
          + \lambda_{np}\, g^{(m)}_{np}
  \label{eq:gtotal}
\end{equation}
where $\lambda_{np}$ is non-negative weighting
factors that trade dc-link balance effort against
tracking accuracy. 
The switching state applied to the inverter is the minimiser of
\eqref{eq:gtotal} over the admissible set,
\begin{equation}
  m^{\star} = \arg\min_{m\in\{1,\dots,N_s\}} g^{(m)}.
  \label{eq:argmin}
\end{equation}
The complete procedure is summarised in Algorithm~1.
 
 \begin{algorithm}[t]
\caption{FCS-MPVC switching-state selection for the 3L-NPC inverter}
\begin{algorithmic}[1]
\Require $\mathbf{x}(k)$, $\mathbf{i}_{\alpha\beta,o}(k)$, $v_{np}(k)$,
         $\mathbf{S}(k-1)$, $\mathbf{v}^{\star}_{\alpha\beta,o}(k+1)$,
         matrices $\mathbf{Q},\mathbf{J},\mathbf{G}$
\Ensure Optimal switching vector $\mathbf{S}^{\star}$
\State $g_{\min} \gets \infty$
\For{$m = 1$ to $N_s = 27$}
  \State Compute candidate voltage $\mathbf{v}^{(m)}_{i,\alpha\beta}$
  \State Predict filter state
         $\mathbf{x}^{(m)}(k{+}1)=\mathbf{Q}\mathbf{x}(k)
          +\mathbf{J}\mathbf{v}^{(m)}_{i,\alpha\beta}
          +\mathbf{G}\mathbf{i}_{\alpha\beta,o}(k)$
  \State Extract $\hat{\mathbf{v}}^{(m)}_{o,\alpha\beta}(k{+}1)
         =\mathbf{C}_v\mathbf{x}^{(m)}(k{+}1)$
  \State Predict midpoint $v^{(m)}_{np}(k{+}1)$ 
  \State Evaluate composite cost $g^{(m)}$ 
  \If{$g^{(m)} < g_{\min}$}
    \State $g_{\min} \gets g^{(m)}$;\quad
           $\mathbf{S}^{\star} \gets \mathbf{S}^{(m)}$
  \EndIf
\EndFor
\State Apply $\mathbf{S}^{\star}$ to the inverter
\end{algorithmic}
\end{algorithm}

 It may appear that, because the controller selects among a finite set of candidate vectors, the problem reduces to rounding an unconstrained continuous solution to the nearest admissible vector. This equivalence holds only when the predicted tracked quantity is an affine, separable function of the applied vector, so that proximity in the input space maps monotonically to proximity in the cost. That is not the case here. The tracked output voltage is obtained through the full coupled state transition of (18), in which the applied vector enters through J while the inherited state enters through Q; the mapping from candidate vector to predicted output is therefore filtered by the second-order LC dynamics and is not a direct scaling of the input. Consequently, the candidate closest to an unconstrained voltage solution is not in general the cost minimiser, and the exhaustive evaluation over the admissible set cannot be replaced by rounding without loss of optimality. The exhaustive search is retained precisely because the coupled model makes the cost non-separable in the input. 
% =====================================================================
%  Section VI — Experimental Results
%  Expanded from the conference paper (linear RL load only)
%  Notation aligned with the journal alpha-beta formulation
% =====================================================================
===============

\section{Simulation Results}
\label{sec:experimental_results}

The proposed FCS-MPVC controller is evaluated through detailed simulations carried out in the Simulink. The objective is to verify that the predictive
formulation derived in Sections~\ref{sec:system_modelling} and
\ref{sec:fcs_mpvc} delivers the expected power quality under the V2L operating conditions.

\subsection{Setup}
\label{sec:setup}

The V2L system of Fig.~\ref{fig:v2l_system} was implemented
in Simulink. The FCS-MPVC algorithm  implemented as a discrete-time
function block executed at the sampling period $T_s$. At every
sampling instant, the algorithm reads the measured state
$\mathbf{x}(k)$, the load current
$\mathbf{i}_{\alpha\beta,o}(k)$, and the reference voltage
$\mathbf{v}_{\alpha\beta,o}^{\star}(k+1)$, evaluates the cost
function for each of the 27 candidate
switching states, and applies the minimizer to the inverter.

\begin{table}[t]
\centering
\caption{Simulation Parameters}
\label{tab:sim_params}
\renewcommand{\arraystretch}{1.15}
\begin{tabular}{l c}
\hline
\textbf{Parameter} & \textbf{Value} \\
\hline
DC-link voltage, $V_{dc}$              & $800$~V \\

EV battery voltage, $V_{\mathrm{bat}}$ & $400$~V \\

Boost inductance, $L_b$                & $0.5$~mH \\

DC-link capacitor 1, $C_1$             & $2000~\mu$F, $100~\mu\Omega$ \\

DC-link capacitor 2, $C_2$             & $2000~\mu$F, $100~\mu\Omega$ \\

Filter inductance, $L_f$               & $5.5$~mH \\

Filter capacitance, $C_f$              & $100~\mu$F \\

Load (active, reactive)                & $10$~kW, $5$~kVAR \\

Sampling period, $T_s$                 & $50~\mu$s \\

\hline
\end{tabular}
\end{table}

\subsection{DC-Link Voltage}
\label{sec:dc_link}
\begin{figure}[!t]
\centering{\includegraphics[width=\columnwidth]{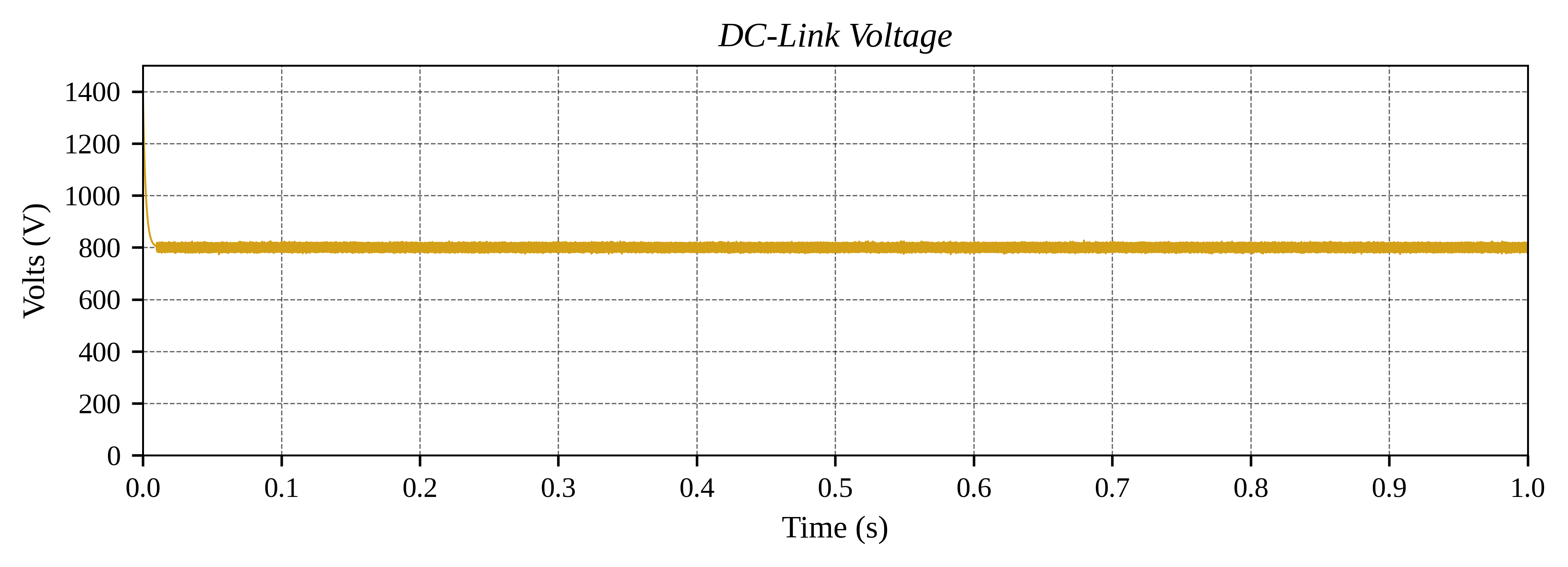}}
\caption{DC-link voltage\label{fig:dc}}
\end{figure}

The boost converter steps up the battery voltage from 400\,V to the 800\,V
dc link required by the inverter and maintains a stable dc-link voltage in
the face of the time-varying current $i_{\mathrm{inv}}$ drawn by the 3L-NPC.
Fig.~\ref{fig:dc} shows the dc-link voltage measured during continuous
operation. The voltage settles rapidly at the 800\, V reference and is
maintained within a peak-to-peak ripple of 1.3\% relative to 800\,V. A
stable, low-ripple dc link is a prerequisite for the predictive controller,
because the candidate inverter voltage vectors scale directly
with $V_{dc}$. Variations in the dc link therefore modulate the entire
candidate set and propagate as voltage error at the output. The 1.3\% ripple
obtained here is sufficiently small to justify the assumption made in
Section~\ref{sec:model} that $V_{dc}$ is approximately constant within one
sampling interval, and validates the use of a fixed dc-link value inside the
predictive model.

\subsection{Controller output}\label{sec:gates}

\begin{figure}[!t]
\centering{\includegraphics[width=\columnwidth]{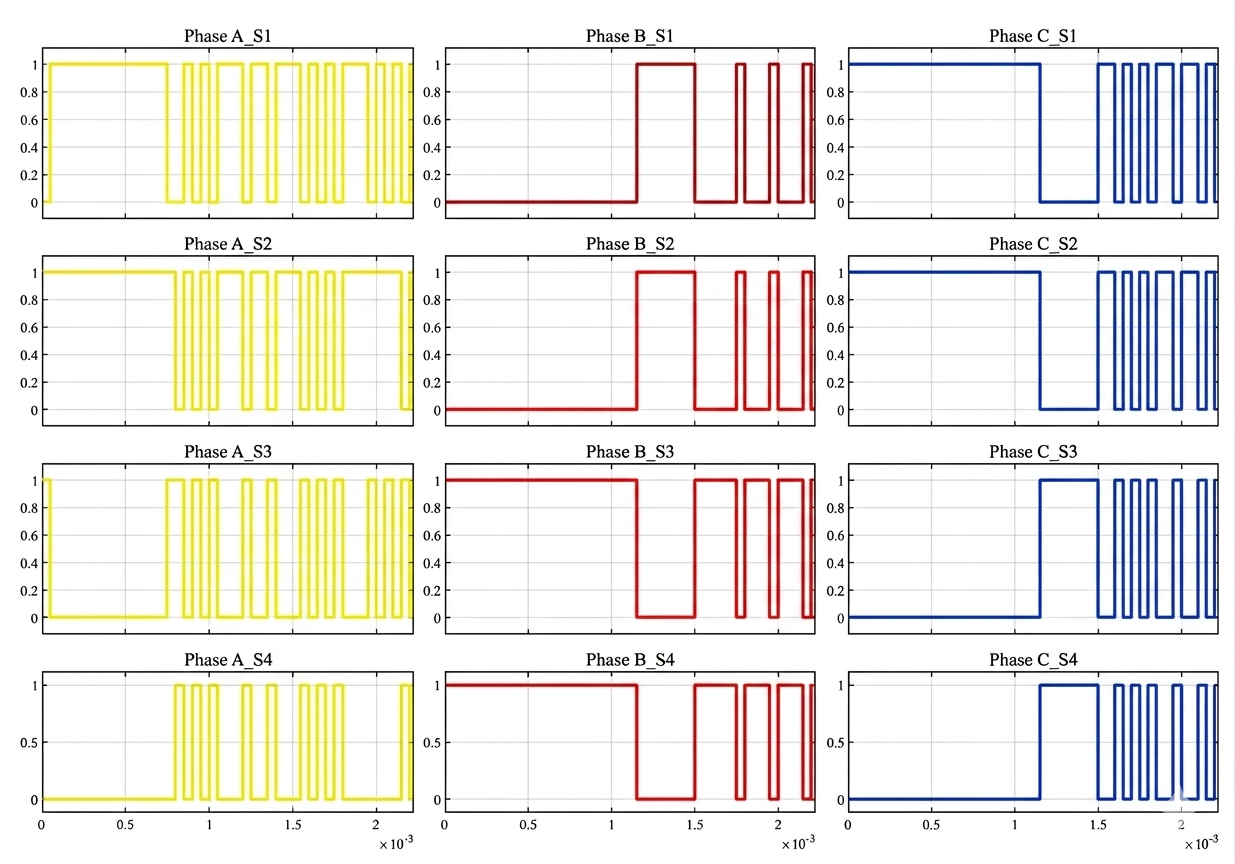}}
\caption{FCS-MPVC switching pulses\label{fig:pulses}}
\end{figure}

Fig.~\ref{fig:pulses} presents the gating signals generated by the FCS-MPVC
for the four switches of each phase leg of the 3L-NPC over a representative
window of operation. The switching transitions occur at irregular intervals
rather than at a fixed period, reflecting the variable switching-frequency
nature of FCS-MPC. At each sampling instant, the controller is free to
repeat the previous switching state or to move to a different one,
depending solely on which candidate minimises the cost. This is
fundamentally different from PWM-based control, where transitions are
clocked by a fixed carrier, and is one of the reasons the FCS-MPVC achieves
low harmonic distortion. Finally, within each phase leg, the four switches
operate in complementary pairs consistent with the 3L-NPC topology. No
short circuit through the dc link is observed, which confirms that the
candidate switching set used by the algorithm represents only
the physically admissible states of the converter.

% If you use the leg-state version of Fig. 4 produced by
% export_iet_figures.m, change the caption to
%   "Leg switching states $S_a$, $S_b$, $S_c$ selected by the FCS-MPVC"
% and in the paragraph above replace "the gating signals generated by the
% FCS-MPVC for the four switches of each phase leg" with "the leg
% switching states $S_x\in\{-1,0,+1\}$ selected by the FCS-MPVC".

\subsection{Steady-state performance}\label{sec:ss}

The steady-state behaviour of the controlled inverter is examined through
the load-side voltage and current waveforms, the voltage tracking error and
the harmonic spectrum of the output.

\subsubsection{Output waveforms}
\begin{figure}[!t]
\centering{\includegraphics[width=\columnwidth]{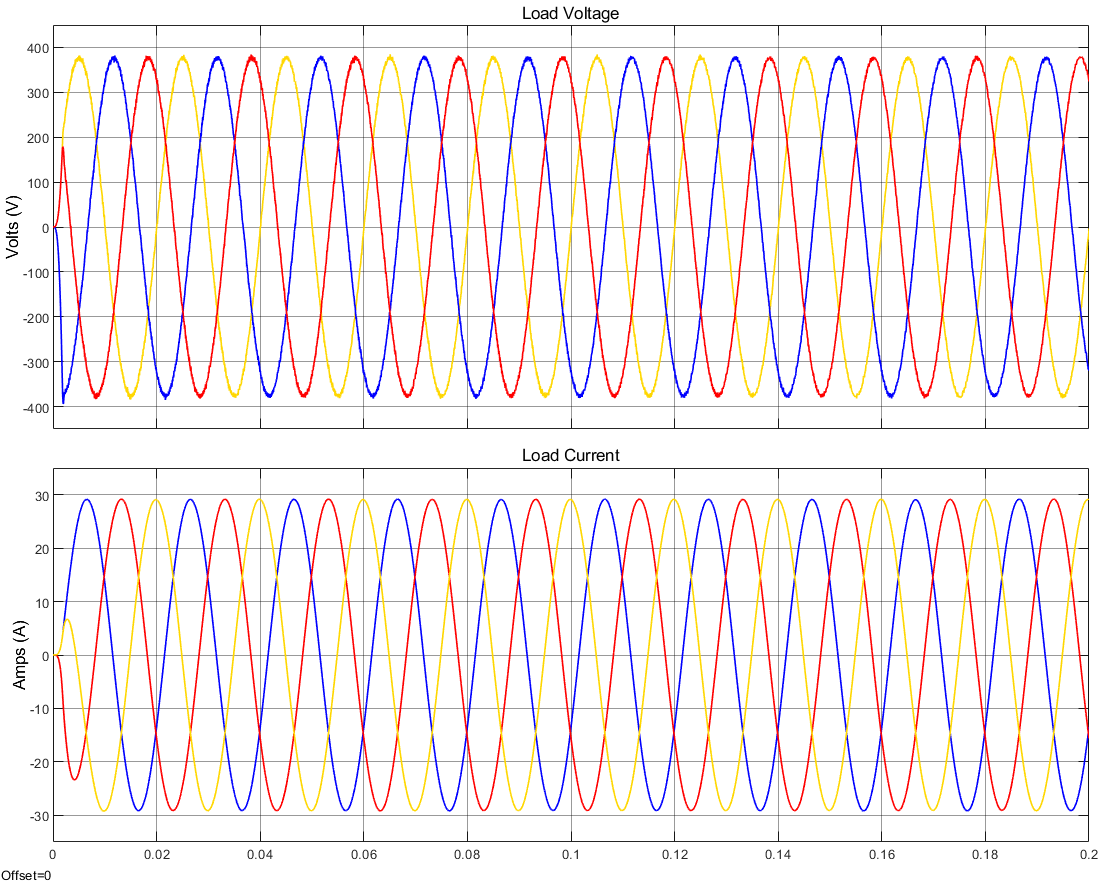}}
\caption{Steady-state load voltage and current\label{fig:ssw}}
\end{figure}

Fig.~\ref{fig:ssw} shows the three-phase load voltage and current under the
nominal load. The voltage waveforms are sinusoidal, balanced and locked to
the reference magnitude and frequency. Start-up from zero initial conditions
is completed quickly, after which the controller operates in steady state
with no observable drift.

\subsubsection{Voltage tracking error}
\begin{figure}[!t]
\centering{\includegraphics[width=\columnwidth]{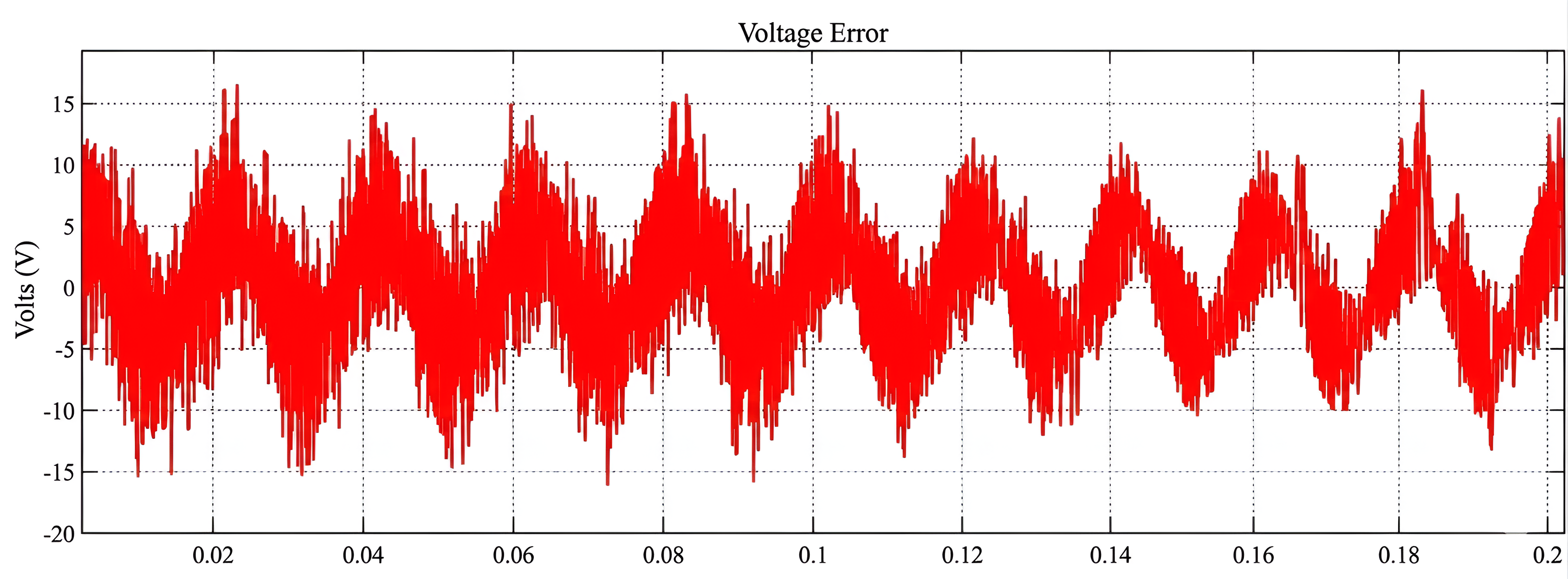}}
\caption{Voltage tracking error\label{fig:err}}
\end{figure}

Fig.~\ref{fig:err} reports the error between the reference and the measured
load voltage during steady state. The root-mean-square (RMS) tracking error
is 1.6\% of the nominal voltage amplitude. Instantaneous excursions reach
about 15\,V and consist mainly of the high-frequency ripple visible in
Fig.~\ref{fig:err}. This residual error has two origins. The first is
the finite size of the candidate set: with only 27 admissible switching
vectors, the controller cannot in general apply the exact reference voltage
at each sampling instant and must select the closest available candidate.
The second is the delay of one sampling period between measurement and
actuation, which is a property of a digital predictive controller.

\subsubsection{Voltage and current harmonic content}
\begin{figure}[!t]
\centering{\includegraphics[width=\columnwidth]{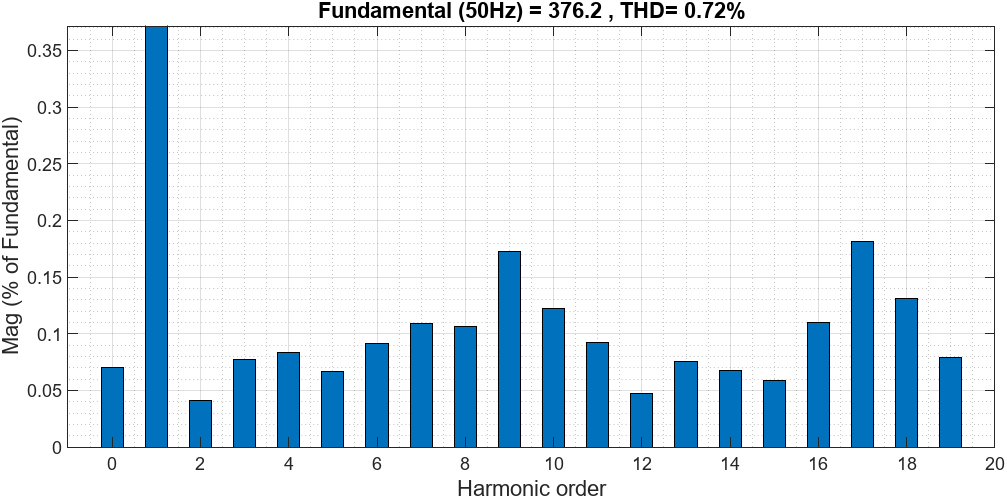}}
\caption{Harmonic spectrum of the load voltage (THD = 0.72\%)\label{fig:vthd}}
\end{figure}

\begin{figure}[!t]
\centering{\includegraphics[width=\columnwidth]{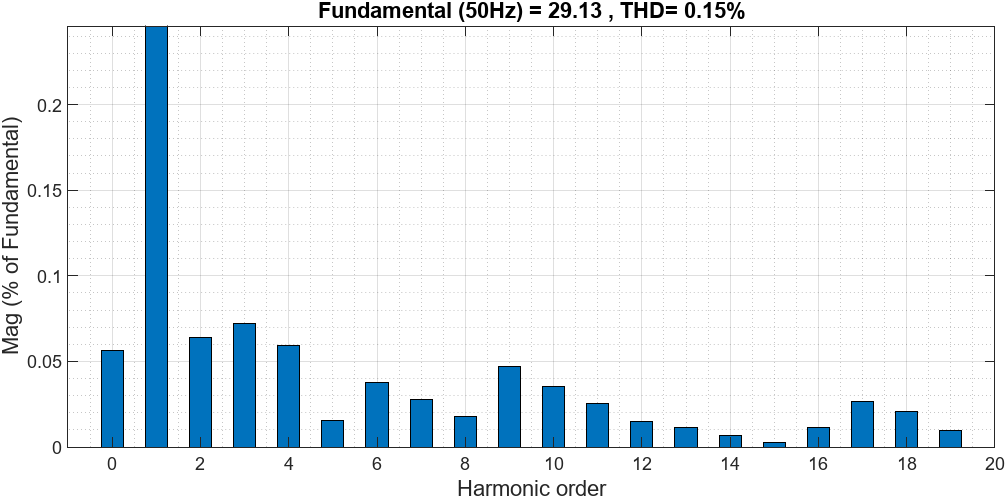}}
\caption{Harmonic spectrum of the load current (THD = 0.15\%)\label{fig:ithd}}
\end{figure}

Fig.~\ref{fig:vthd} shows the harmonic spectrum of the load voltage referred
to the fundamental. The total harmonic distortion (THD) is 0.72\%, with the
residual content concentrated in the upper end of the analysed band,
reflecting the high attenuation provided by the LC filter at frequencies
well above its 215\,Hz cut-off. For completeness, Fig.~\ref{fig:ithd}
presents the harmonic spectrum of the load current, which exhibits a still
lower THD of 0.15\%.

\subsubsection{Neutral-point voltage}
\begin{figure}[!t]
\centering{\includegraphics[width=\columnwidth]{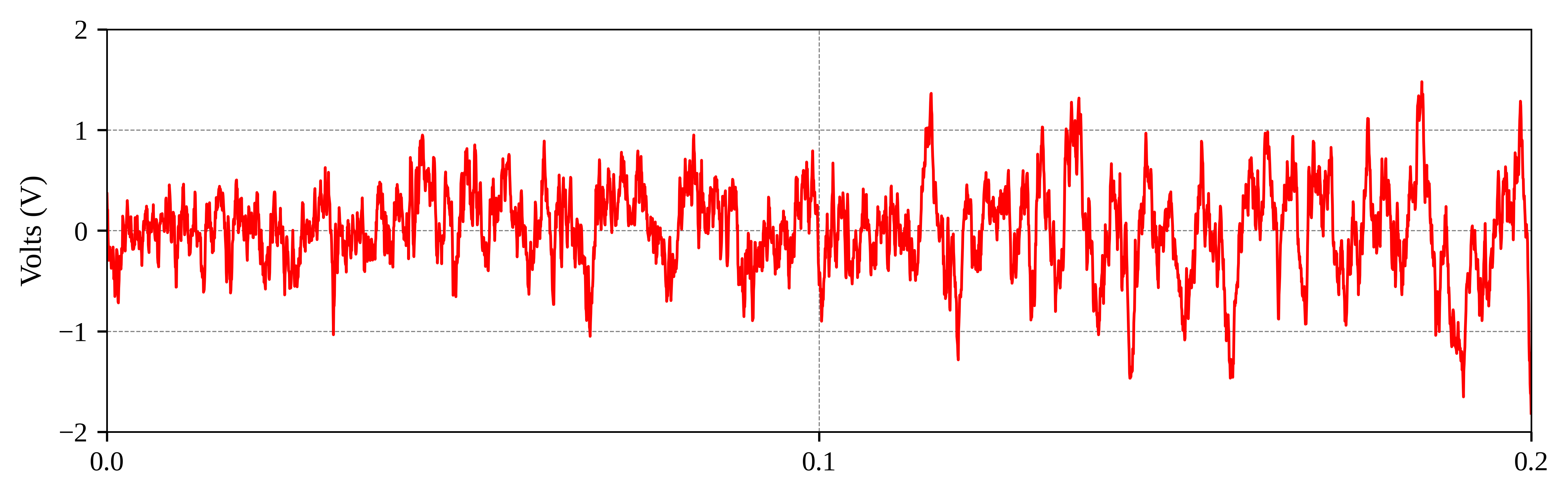}}
\caption{Neutral-point voltage\label{fig:np}}
\end{figure}

Fig.~\ref{fig:np} shows the neutral-point voltage of the dc link during
steady-state operation. With the neutral-point term included in the cost
function, the relative voltage imbalance is low. This behaviour follows the
mechanism analysed in Section~\ref{sec:np}: among switching states that
produce similar predicted output voltages, the controller exploits the
redundancy of the 3L-NPC and selects the one that minimises the cost
function.

\subsection{Transient performance}\label{sec:tr}

The transient performance of the controller is assessed under a step
doubling of the load active power applied at $t=0.1$\,s. This represents a
severe disturbance scenario: the load demand changes abruptly, the load
current must follow without prior knowledge of the event, and the controller
must accommodate the new operating point on the basis of the
receding-horizon update alone.

\begin{figure}[!t]
\centering{\includegraphics[width=\columnwidth]{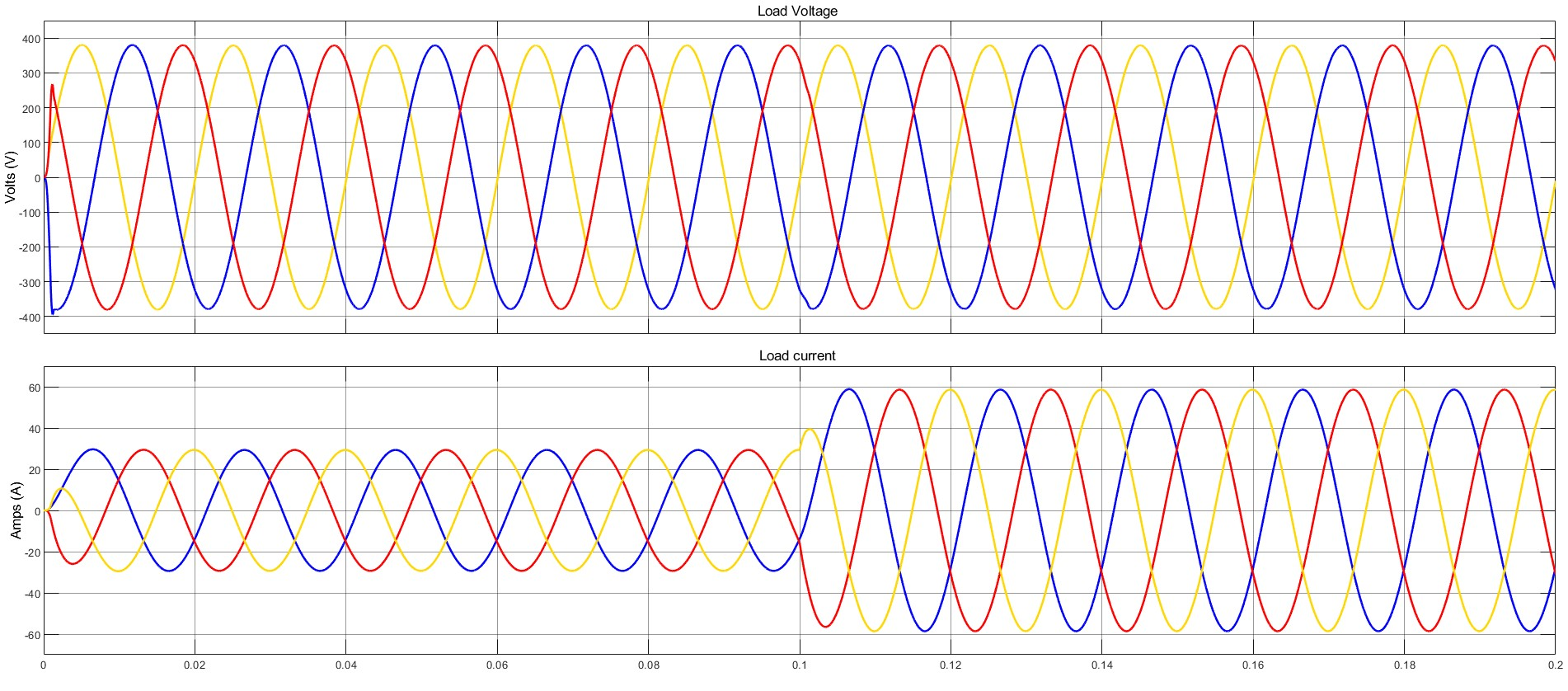}}
\caption{Transient waveforms: load voltage and current under a step doubling of the load active power at $t=0.1$\,s\label{fig:tr}}
\end{figure}

As shown in Fig.~\ref{fig:tr}, the load voltage tracks the reference
throughout the disturbance. The voltage envelope remains continuous and
phase-locked to the reference. In the current waveform, a peak overshoot of
3\% and a settling time of 15\,ms are observed. The settling time is shorter
than the 20\,ms period of the 50\,Hz fundamental, so steady-state operation
is re-established within less than one cycle of the output voltage. The peak
overshoot is consistent with the natural second-order response of the LC
filter.

\subsection{Comparison with power-quality limits}\label{sec:ieee}

To place the obtained results in context, Table III compares the
steady-state and transient metrics of the proposed FCS-MPVC with limits
taken from the relevant power-quality standards. The voltage THD of 0.72\%
is about eleven times below the 8\% limit that IEEE Std 519 sets for buses
at or below 1\,kV [35]. The load-current THD of 0.15\% is more
than thirty times below both the 5\% total demand distortion limit of the
same standard for the most restrictive system class
($I_{sc}/I_L<20$) and the 5\% total rated current
distortion limit that IEEE Std 1547 applies to distributed energy
resources [35]. The RMS tracking error of 1.6\% bounds the
deviation of the fundamental output voltage: since the fundamental
component of the error cannot carry more energy than the whole error, its
amplitude is at most $\sqrt{2}\times1.6\%\approx2.3\%$ of the nominal
amplitude, well inside the $\pm5\%$ service-voltage band. No standard
directly prescribes the transient overshoot and recovery time of an
islanded V2L source. For these two metrics, the design targets adopted in this work are an overshoot below
5\% and recovery within one fundamental cycle (20\,ms), both of which are
met. Taken together, these results indicate that the proposed FCS-MPVC, in its
fundamental form, meets the applicable power-quality requirements without
observers, parameter estimation or learning-based augmentation.

\subsection{Comparison with recent predictive controllers}\label{sec:lit}

The finite-set controllers in [36] and [39] achieve output-voltage
THDs of 1.15--2.92\%, while the CCS-MPC of [37] reaches 1.28\% by
combining a modulator with a disturbance observer. The proposed controller
achieves 0.72\% while retaining the single-vector, modulator-free structure
of conventional FCS-MPC, without an observer. This result also benefits
from the three-level topology and the filter design adopted here.

The robustness studies by authors of [37] and [28] show that model
mismatch is the main cause of degradation in conventional predictive
controllers, with the THD rising to 5.81\% and 7.8\% under parameter
errors. This supports the emphasis on model fidelity in the present work. In addition, the
neutral-point voltage remains within approximately
$\pm2$\,V on an 800\,V dc link, compared with deviations of up to 2\,V on
a 100\,V dc link reported in [40].

% =====================================================================

\subsection{Prediction-Accuracy Bound}
\label{sec:prediction_accuracy}

The accuracy with which (18) reproduces the actual filter
response over one sampling interval can be bounded analytically. Four
potential sources of one-step prediction error are distinguished below:
disturbance-truncation error, dc-link voltage error, discretisation
error, and parametric error. The
analysis is carried out at the parameters of the simulated
system, namely $L_f = 5.5$~mH, $C_f = 100~\mu$F, and
$T_s = 50~\mu$s, which give a filter natural frequency
\begin{equation}
\omega_n \;=\; \frac{1}{\sqrt{L_f C_f}}
\;\approx\; 1{,}348~\mathrm{rad/s}\;\;(\approx\!215~\mathrm{Hz}).
\label{eq:omega_n_value}
\end{equation}

\subsubsection{Disturbance-Truncation Error}
The load current $\mathbf{i}_{\alpha\beta,o}(\tau)$ varies
during $[kT_s,\,(k+1)T_s]$ but the predictive model uses its
sampled value $\mathbf{i}_{\alpha\beta,o}(k)$ throughout the
interval. The resulting one-step prediction error is:
\begin{equation}
\bm{e}(k\!+\!1) \;=\;
\int_{0}^{T_s} e^{\mathbf{A}(T_s-\tau)}\,\mathbf{E}\,
\big[\mathbf{i}_{\alpha\beta,o}(kT_s\!+\!\tau)
- \mathbf{i}_{\alpha\beta,o}(kT_s)\big]\,d\tau
\label{eq:trunc_err}
\end{equation}
For a sinusoidal load current of fundamental frequency
$f_1 = 50$~Hz and peak amplitude $\hat{I}_o$, the integrand is
bounded above by $2\pi f_1\,\hat{I}_o\,\tau$, and standard
integration of \eqref{eq:trunc_err} yields the tight bound
\begin{equation}
\|\bm{e}(k+1)\|_2 \lesssim \frac{T_s^2}{2}\,\|\bm{E}\|_2\,(2\pi f_1 \hat{I}_o),
\label{eq:trunc_bound}
\end{equation}
With $\|\mathbf{E}\|_{\,2} = 1/C_f = 10^{4}~\text{V/As}$,
$T_s = 50~\mu$s, $f_1 = 50$~Hz, and a representative peak load
current $\hat{I}_o = 30$~A, the bound evaluates to
approximately $118$~mV, equivalent to $0.04\%$ of the nominal
phase-voltage peak. The disturbance-truncation contribution to
the one-step prediction error is therefore negligible at the
chosen sampling rate.

\subsubsection{DC-link voltage error}
The dc-link voltage $V_{dc}$ appears in the candidate inverter voltages
$\mathbf{v}_{\alpha\beta}^{(m)}$ through
\eqref{eq:vi_alphabeta}. Variations in $V_{dc}$ rescale the
entire candidate set, but because the optimisation in
\eqref{eq:argmin} is evaluated against the same reference
$\mathbf{v}_{\alpha\beta,o}^{\star}(k+1)$, the controller
naturally selects whichever (rescaled) candidate is closest to
that reference. The $1.3\%$ ripple on $V_{dc}$ demonstrated in
Section~\ref{sec:dc_link} is well within the margin where this
self-correction is effective, and no performance degradation is
observed in the steady-state results of
Section~\ref{sec:steady_state}.
\subsubsection{Discretisation Error}
The discrete-time model \eqref{eq:disc_ss} is obtained from the
exact zero-order-hold discretisation \eqref{eq:zoh_matrices},
so that under the assumptions that (i) the inverter voltage is
constant over $[kT_s,\,(k+1)T_s]$ and (ii) the load current is
also constant over the same interval, the
prediction~\eqref{eq:state_pred} is exact. Assumption (i) is
satisfied by construction: the FCS-MPVC selects a single
switching state $\mathbf{S}^{(m^{\star})}$ at instant $kT_s$
and holds it until the next sampling instant. Assumption (ii)
is approximate, and is analysed next.

\subsubsection{Parametric Error}
The 4th source is parametric: any deviation between the
values of $L_f$ and $C_f$ used in $\mathbf{Q}$, $\mathbf{J}$,
$\mathbf{G}$ and the actual hardware values appears as a
multiplicative error on the predicted state. In practice these
parameters may deviate from their nominal values due to
manufacturing tolerances, temperature, and ageing. Their
combined effect can be quantified directly through the filter
natural frequency $\omega_n$, since the per-sample prediction
error is of order $\omega_n T_s$. At the nominal parameters,
\begin{equation}
\omega_n T_s \;\approx\; 0.067~\mathrm{rad}
\;\approx\; 3.9^{\circ}\text{ per sample}.
\label{eq:phase_error_value}
\end{equation}
A predictive controller built on a decoupled filter model
would accumulate this $3.9^{\circ}$ phase error between its
prediction and the actual evolution of
$\mathbf{v}_{\alpha\beta,o}$. At a $50$~Hz fundamental, this
corresponds to a steady-state voltage error of approximately
$\sin(3.9^{\circ}) \approx 6.8\%$ of the fundamental amplitude,
an order of magnitude larger than the $1.6\%$ measured with the
proposed coupled formulation
(Section~\ref{sec:steady_state}).

Parametric mismatch shifts $\omega_n$ and therefore shifts the
per-sample phase error. Differentiating $\omega_n$ with respect
to $L_f$ and $C_f$ at the nominal operating point yields
\begin{equation}
\frac{\Delta \omega_n}{\omega_n}
\;\approx\; -\tfrac{1}{2}\frac{\Delta L_f}{L_f}
\;-\; \tfrac{1}{2}\frac{\Delta C_f}{C_f},
\label{eq:omega_sens}
\end{equation}
so that the natural frequency varies at half the rate of the
parameter deviation. Table~\ref{tab:sensitivity} reports the
corresponding shifts for component tolerances of $\pm 10\%$
and $\pm 20\%$ on $L_f$, with $C_f$ held at its nominal value.

\begin{table}[t]
\centering
\caption{Sensitivity of the Predictive Model to Filter-Inductance Variation}
\label{tab:sensitivity}
\renewcommand{\arraystretch}{1.15}
\begin{tabular}{c c c c}
\hline
$\Delta L_f / L_f$ & $\omega_n$ (rad/s) & $\Delta \omega_n / \omega_n$
                                        & Phase error / sample \\
\hline
$-20\%$ & $1{,}508$ & $+11.8\%$ & $4.3^{\circ}$ \\
$-10\%$ & $1{,}421$ & $\phantom{+}{+}5.4\%$  & $4.1^{\circ}$ \\
nominal & $1{,}348$ & $\phantom{+}0$         & $3.9^{\circ}$ \\
$+10\%$ & $1{,}286$ & $-4.7\%$  & $3.7^{\circ}$ \\
$+20\%$ & $1{,}231$ & $-8.7\%$  & $3.5^{\circ}$ \\
\hline
\end{tabular}
\end{table}

Across the full $\pm 20\%$ range, the per-sample phase error
changes by no more than $\pm 0.4^{\circ}$, which is small in
absolute terms and is corrected within a single sampling
interval by the receding-horizon update. At each step $kT_s$,
the controller measures the actual state $\mathbf{x}(k)$ and
the actual load current $\mathbf{i}_{\alpha\beta,o}(k)$, then
re-solves the optimisation \eqref{eq:argmin} from the true
operating point. Prediction errors arising from parameter
mismatch therefore do not accumulate over time but are reset at
every sample.

% ---------------------------------------------------------------------
% ITEM 5 — computational complexity with concrete cycle counts
% Replaces equations (24)-(26) and the surrounding text in your
% current Section V-A.
% ---------------------------------------------------------------------
\section{IMPLEMENTATION CONSIDERATIONS}

\subsection{Computational Complexity}
\label{sec:complexity}

A central advantage of FCS-MPC is that the optimal switching state is selected
directly by evaluating the predicted behaviour for every
admissible switching combination. At every
sampling instant, the controller performs three operations
for each candidate: (i) one matrix-vector evaluation of the
discrete-time prediction~\eqref{eq:state_pred}; (ii) one
evaluation of the voltage-tracking cost and the
neutral-point cost; and (iii) one comparison
against the running minimum.

Step (i) requires multiplication of the state-transition
matrix $\mathbf{Q} \in \mathbb{R}^{n\times n}$ (with $n=4$) by
the state $\mathbf{x}(k)$, and similar products of
$\mathbf{J}$ and $\mathbf{G}$ with the candidate input and the
measured load current. The asymptotic cost per candidate is
$\mathcal{O}(n^2)$, giving a total per-step cost of
\begin{equation}
\mathcal{O}(N_s\,n^2) \;=\; \mathcal{O}(432).
\label{eq:complexity_bound}
\end{equation}

Including the voltage-tracking cost \eqref{eq:gv} and the neutral-point
prediction \eqref{eq:vnp-pred}, the realistic operation count is approximately
$2\times10^{3}$ floating-point operations per sampling step, equivalent
to a throughput of 40~MFLOPS at $T_s = 50~\mu$s.

\subsection{Implementation and scalability}
Although the proposed control strategy is evaluated through
simulation, its structure is compatible with real-time
implementation in practical inverter systems.

\subsubsection{Measurement Requirements}
The controller requires measurements of the output voltages
and filter currents to construct the state vector used for
prediction. These are obtained through sensors connected to
high-speed analog-to-digital converters; synchronisation
between measurements and control updates is essential to
maintain prediction accuracy.

\subsubsection{Sampling Time Selection}
A smaller sampling period $T_s$ improves prediction accuracy
and dynamic response at the cost of higher computational
demand. The value $T_s = 50~\mu$s adopted here represents a
compromise between control bandwidth and the throughput
requirement quantified in Section~\ref{sec:complexity}.

\subsubsection{Switching-Frequency Behaviour}
FCS-MPC produces a variable switching frequency, since the
optimal state is selected at each sampling instant rather
than at a fixed carrier rate. This contributes to fast
transient response but complicates electromagnetic-compatibility
analysis and filter design. Practical implementations may
add cost-function terms or constraints to regulate switching
transitions.

\subsubsection{Digital Controller Implementation}
The algorithm is suited to implementation on standard DSPs or
FPGAs. Because the control decision involves a finite number
of candidates and no iterative solver, the required
computations complete within the sampling interval, and the
software implementation is straightforward.

\subsection{Measurement Error and Sensor Faults}
The controller reconstructs the state from measured voltages and currents at each step; measurement error therefore enters directly through x(k). Because the receding-horizon update re-anchors the prediction to the fresh measurement at every sample rather than integrating past estimates, a bounded measurement perturbation produces a bounded, non-accumulating one-step prediction error, of the same order as the disturbance-truncation term analysed above. However, the amplification of broadband measurement noise into switching activity, and tolerance to partial sensor loss, are not evaluated in this study and would require either measurement filtering or an observer-based reconstruction; both are left to the experimental phase of this work.

\section{Discussion}
\label{sec:discussion}

The results of Section~\ref{sec:experimental_results} confirm
the predictions of the model developed in
Sections~\ref{sec:system_modelling} and~\ref{sec:fcs_mpvc}.
The voltage tracking error and THD measured at
the load terminals are very low. The transient response to a doubling of the load recovers within less than one fundamental cycle, with an overshoot consistent with the natural damping of the $LC$ filter, and the voltage remains on track throughout the transient. The neutral-point
voltage is held within a small fraction of the dc-link.

The receding-horizon update prevents prediction
errors from accumulating across sampling instants by
re-anchoring the prediction to the measured state
$\mathbf{x}(k)$ at every step. The analytical bound derived in
Section~\ref{sec:prediction_accuracy} is consistent with the
$1.6\%$ steady-state voltage error observed empirically, the
residual being attributable to the finite size of the
candidate set. The results suggest that substantial performance can be achieved without parameter estimation, disturbance observers, or machine-learning-assisted augmentation when sufficient predictive model fidelity is retained.

The results reported here are obtained entirely in simulation. Although the discrete-time model, the candidate evaluation, and the throughput analysis are constructed to reflect a realizable fixed-point implementation, effects that a full-hardware or controller-hardware-in-the-loop study would expose require a controller-hardware-in-the-loop and experimental validation, the primary direction of ongoing work.

\section{Conclusion}
This paper presented a coupled-state FCS model predictive voltage controller for a 3L-NPC inverter in vehicle-to-load operation. By carrying the filter current and output voltage jointly in the predictive model, the controller retains the full second-order LC coupling without enlarging the candidate set, while the cost function additionally maintains neutral-point balance. Simulation results show low voltage THD, small tracking error, and transient recovery within one fundamental cycle, with closed-form bounds confirming low sensitivity to load variation and parameter mismatch. The results indicate that predictive-model fidelity alone can deliver high power quality, reducing the need for observers or learning-based augmentation. Ongoing work targets controller-hardware-in-the-loop and experimental validation, and a mode selector for seamless V2L, V2G, and G2V transitions.

%-----------------------

% or


% ============================================
%\appendices
%Appendix one text goes here %\cite{Roberg2010}.

%\section{}

%\section*{Acknowledgment}

%The authors would like to thank D. Root for the loan of the SWAP. The SWAP that can ONLY be usefull in Boulder...

\ifCLASSOPTIONcaptionsoff
  \newpage
\fi

\vfill


\end{document}